\documentclass{emulateapj}

\shorttitle{Forbidden Lines in AGNs}
\shortauthors{Devereux et al.}

\begin{document}

\title{On the Absence of Broad Forbidden Emission Lines in the Low Luminosity Seyfert 1 Nucleus of NGC 3227}

\author{Nick Devereux}
\affil{Department of Physics, Embry-Riddle Aeronautical University,
         Prescott, AZ 86301}
\email{devereux@erau.edu}

\begin{abstract}
The absence of broad [O\,{\sc iii}] ${\lambda\lambda}$4959,5007 forbidden emission lines is one of the key arguments cited in the published literature for the gas density exceeding the critical density; 7 ${\times}$ 10$^5$ cm$^{-3}$, in the broad line region (BLR) of active galactic nuclei. However, for NGC 3227, an equally valid alternative explanation is that O$^{2+}$ is progressively ionized to O$^{3+}$ as the central UV--X-ray source is approached. Observational evidence for such an ionization gradient is provided by spectra obtained with the {\it Space Telescope Imaging Spectrograph}. Modeling the rich UV--visible spectrum of NGC 3227 with the photoionization code Cloudy supports the interpretation that the absence of broad [O\,{\sc iii}] forbidden emission lines is due to ionization, rather than high gas density, and further suggests that what we perceive as the BLR in NGC 3227 is just the illuminated portion of a much larger inflow. The low metallicity deduced for the inflowing gas suggests an origin in the circumgalactic medium.
\end{abstract}


\section{Prologue} 
This paper was rejected from two prestigious journals.  Just as well as the Cloudy modeling results reported herein are highly uncertain because Cloudy has a bug in it. I had not discovered the bug until after I had written \& submitted this manuscript. The bug is described in more detail in the forthcoming ApJ paper
http://adsabs.harvard.edu/abs/2015arXiv151204604D
On a positive note, the observational results reported in this paper are correct, in particular the HST emission line fluxes. I think the reason that the manuscript was rejected is because of the idea expressed that photoionization, rather than gas density, is the reason for the absence of broad forbidden emission lines. Evidently, this possibility touched a raw nerve among the reviewers. However, that idea is expressed again
in the aforementioned ApJ paper that will be published in 2016.

\section{Introduction}

A basic tenet with regard to the astrophysics of active galactic nuclei (AGN) is that the Hydrogen (H) gas number density in the broad line region (BLR) must exceed
10$^6$ cm$^{-3}$ because of the absence of broad forbidden lines. The genesis of this widely held sentiment
can be traced to \cite{KW71}, and \cite{S72}, who argued that `high density rather than high ionization is responsible for the absence of broad components to the forbidden lines' in the Seyfert 3C 120.  Specifically, 
the gas number density in the BLR must exceed the critical density of 7 ${\times}$ 10$^5$ cm$^{-3}$
for collisional de-excitation of the ${^1D_2}$ level that would otherwise produce broad [O\,{\sc iii}]${\lambda\lambda}$4959,5007 forbidden emission lines\footnotemark \footnotetext[1]{Interestingly, the critical density
is higher, 2 ${\times}$ 10$^7$ cm$^{-3}$, for the ${^1S_0}$ level responsible for the [O\,{\sc iii}]${\lambda}$4363 forbidden emission line \citep{Rub89} but, unfortunately, this line is weak in NGC 3227 and the line profile shape is overwhelmed by emission from the red wing of the broad H${\gamma}$ line \cite[see][for details]{D13}.}.  

However, in the case of NGC 3227, a contrary argument can be made that 
the O$^{2+}$ ions required for the production of broad [O\,{\sc iii}] forbidden emission lines are 
absent because they have been ionized to O$^{3+}$. This reasoning follows from the observation that
broad permitted C\,{\sc iv} ${\lambda}$1549 line emission is ubiquitous in AGNs including NGC 3227 \cite[][and references therein]{S07}. Permitted C\,{\sc iv} ${\lambda}$1549 line emission results from collisionally excited C$^{3+}$
ions which implies photons with energies exceeding the 48 eV required to ionize C$^{2+}$. With a similar ionizing potential of ${\sim}$ 55 eV, O$^{2+}$ would also be ionized to O$^{3+}$ yielding  [O\,{\sc iv}]  line emission. Indeed, a bright [O\,{\sc iv}] emission line is seen in NGC 3227 at 25.9 ${\mu}$m as reported by
\cite{D11} although the limited angular resolution of the {\it Spitzer Space Telescope} prevents a definitive measurement of any BLR component. Nevertheless, it is conceivable that the coincidence of broad C\,{\sc iv} emission and the absence of broad  [O\,{\sc iii}] emission is due to high ionization rather than high density, contrary to widely held belief. To be fair, in defending collisional de-excitation as the explanation for the absence of broad  [O\,{\sc iii}] emission lines in the Seyfert 3C 120, \cite{S72} did acknowledge the caveat `unless X-ray ionization is important'. With hindsight we now know that broad line AGNs produce X-rays \citep[][and references therein]{Win11}. Thus, photoionization of O$^{2+}$ may well be the reason behind the absence of broad [O\,{\sc iii}] forbidden emission lines in NGC 3227, and, perhaps, other AGNs.

The purpose of this {\it Paper} is to report a new observational result, described in more detail in Section 2, which suggests that an ionization gradient exists in the central 1pc of NGC 3227.  The new result is based on UV--visible spectroscopy obtained with the {\it Hubble Space Telescope (HST)}. As discussed in Section 3, modeling the spectra using the photoionization code, Cloudy \citep{F13}, supports the interpretation that the absence of broad  [O\,{\sc iii}] emission lines is due to ionization, not high density, at least in NGC 3227.  Conclusions are presented in Section 4.

\section{Results}

The exquisite UV-visible spectra of the Seyfert 1.5 nucleus in NGC 3227 obtained with the Space Telescope Imaging Spectrograph ${\it (STIS)}$
aboard ${\it HST}$ have been presented most recently by \cite{D13} and will not be discussed further here other than to say that Table 1 summarizes additional information for UV emission lines seen with the G140L and G230L gratings. Specifically, the vacuum wavelength Mg\,{\sc ii} ${\lambda}$2798,  C\,{\sc iii}] ${\lambda}$1908, He\,{\sc ii} ${\lambda}$1640 and C\,{\sc iv} ${\lambda}$1542 emission lines.

\begin{deluxetable}{cccc}
\tabletypesize{\scriptsize}
\tablecaption{Emission Line Parameters for the G140L and G230L Nuclear Spectra Obtained 2--8--2000\tablenotemark{a}}
\tablewidth{0pt}
\tablehead{
\colhead{Line} & \colhead{Central Wavelength\tablenotemark{b}} & \colhead{Flux\tablenotemark{c}} & \colhead{FWHM}   \\
\colhead{} & \colhead{\AA} &  \colhead{10$^{-14}$ erg cm$^{-2}$ s$^{-1}$} & \colhead{kms$^{-1}$} \\
\colhead{(1)} & \colhead{(2)} &  \colhead{(3)} & \colhead{(4)} \\
}
\startdata
$\textrm{C IV}$\tablenotemark{d} &  1555  ${\pm}$ 1 &    ${\geq}$ 7.4  & 4000  \\
$\textrm{C III]}$ &  1913  ${\pm}$ 1 &   9.0 ${\pm}$  0.3 & 4730 ${\pm}$ 207 \\
$\textrm{ Mg II}$\tablenotemark{d} &  2810 ${\pm}$ 3 &  ${\geq}$ 11.5 &  2750 \\
$\textrm{He II}$ &  1645  ${\pm}$ 1 &   1.4 ${\pm}$  0.1 & 3070 ${\pm}$ 337 \\

\enddata
\tablenotetext{a}{Table entries that do not include uncertainties are fixed parameters. }
\tablenotetext{b}{Observed wavelength}
\tablenotetext{c}{Measured within
a 0.2{\arcsec}  x 0.35{\arcsec}  aperture. Continuum subtracted but not corrected for dust extinction. Model dependent systematic uncertainties introduce an additional ${\sim}$3\% error not reported in the Table.}
\tablenotetext{d} {The flux is a lower limit because emission line exhibits absorption features.}

\end{deluxetable}

The evidence for an ionization gradient is presented in Figure 1 which illustrates a correlation between FWHM and ionization potential, separately for permitted, and forbidden lines in NGC 3227. The permitted lines show a similar range in ionization potential as the forbidden lines, although the ionization gradient for the forbidden lines is steeper. The significant velocity offset between the permitted and forbidden lines illustrates the conspicuous absence of broad forbidden lines in NGC 3227.
In contrast to the correlation presented in Figure 1, there is no significant correlation between the FWHM and critical density for forbidden lines as illustrated in Figure 2.

\begin{figure}
 \includegraphics[width=84mm,clip]{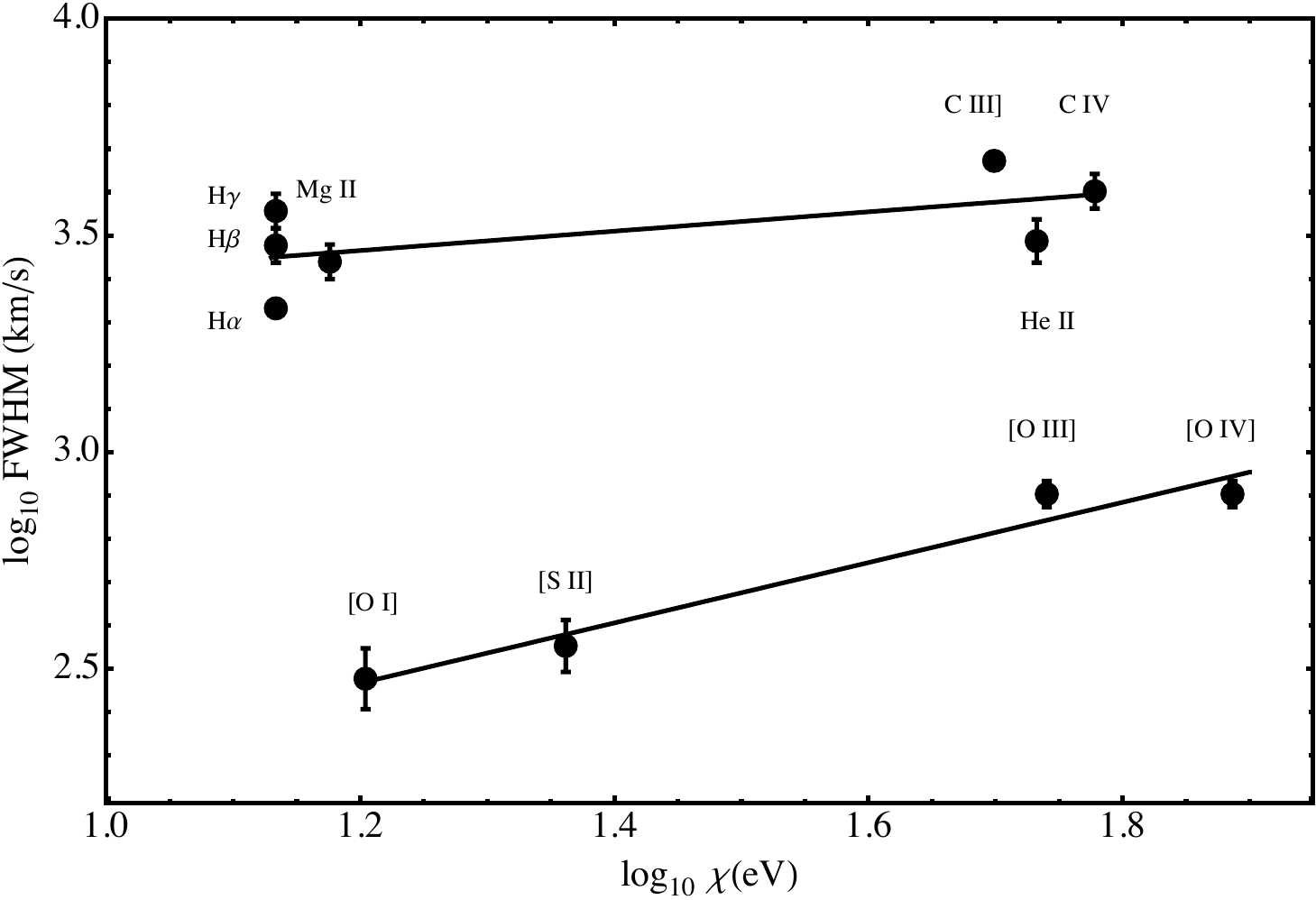}
 \caption{Correlation between FWHM and ionization potential for permitted, and forbidden emission lines in NGC 3227. Following \cite{D11}, the [O\,{\sc iv}] data point
has been included with a FWHM set equal to that of the [O\,{\sc iii}] line.  }
\end{figure}

\begin{figure}
 \includegraphics[width=84mm,clip]{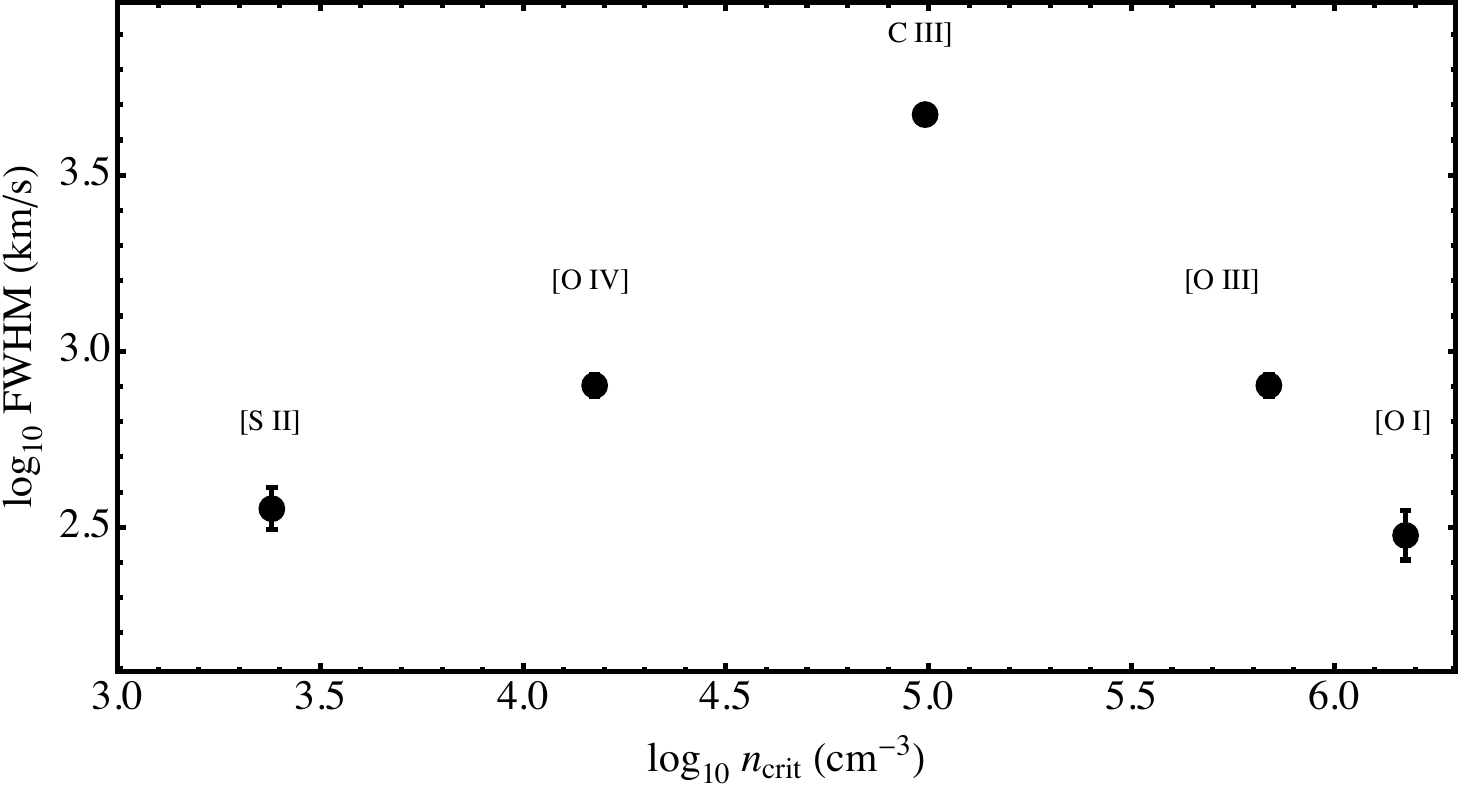}
 \caption{No correlation between FWHM and critical density for forbidden lines in NGC 3227.  Following \cite{D11}, the [O\,{\sc iv}] data point has been included with a FWHM set equal to that of the [O\,{\sc iii}] line. }
\end{figure}

\vspace{1.0cm}

\section{Discussion}
An ionization gradient is expected in the nebula surrounding the central UV--X-ray source in NGC 3227 for the same reasons that one exists in the nebula surrounding H{\sc ii} regions 
and planetaries \citep{Ost06}. However, the exact nature of the ionization gradient depends on the ionizing spectrum of the central UV--X-ray source and the density of the surrounding gas all of which has been explored using version 13.02 of the photoionization modeling code, Cloudy \citep{F13}. Specific parameters employed to model the photoionization in NGC 3227 are listed in Table 2,
and describe a spherically symmetric steady state inflow that is photoionized by the central UV--X-ray source. Following \cite{Vas09}, the shape of the ionizing spectrum is represented by the superposition of a 10$^5$ K blackbody, and a power law with spectral index ${\alpha_{\rm{ox}}}$=-1.17. The number of ionizing photons, q(h), is constrained to be 52.8 dex \citep{D13}. A spherical volume of neutral gas is described by an inner, and outer radius, and a radial density profile constrained by the requirement for a steady state inflow.
The main attribute of the inflow model is that it can easily reproduce the shape of the single peak broad emission lines seen in NGC 3227 \citep{D13}.
Dust grain sublimation is also included. A full description of the parameters can be found in the Cloudy documentation.

\begin{deluxetable}{l}
\tabletypesize{\scriptsize}
\tablecaption{Input Parameters for the Cloudy Photoionization model of NGC 3227}
\tablewidth{0pt}
\tablehead{
\colhead{Parameter}   \\
}
\startdata
AGN T=1e5 K, $\alpha_{\rm{ox}}$=-1.17 \\
q(h)=52.8 \\
radius 15.0 20.0 \\
hden 8.95, power =-1.5 \\
sphere \\
aperture slit \\
abundances ISM no grains \\
grains function sublimation \\
element scale factor carbon 0.1 \\
element scale factor nitrogen 0.1 \\
element scale factor oxygen 0.1 \\
Stop temperature 2 \\
iterations 2 \\
\enddata
\end{deluxetable}

\subsection{Ionization Structure of the Broad and Narrow Lines Regions. }

Photoionization modeling of the inflow fueling the AGN in NGC 3227 yields the ionization structure illustrated in Figure 3 which was generated by Cloudy using the parameters listed in Table 2. The principal merits of this particular model are that the size of the H$^+$ region that is completely ionized coincides with the size of the BLR inferred from the shape of the broad H${\alpha}$ emission line profile \cite[see][]{D13}, and identified in Figure 3 by the vertical dashed lines. Also, the ionization fraction progressively gives way to higher stages of ionization up to, and including O$^{3+}$, as the central UV--X-ray source is approached
suggesting an ionization gradient akin to the one observed, and illustrated in Figure 1. 
Whilst only a proxy, the similar extent of the O$^{2+}$ and O$^{3+}$ regions may explain the similarity between the  [O\,{\sc iii}] and [O\,{\sc iv}] emission line profiles noted previously by \cite{D11}. 
However, since both lines are forbidden, they are susceptible to collisional de-excitation.

Cloudy predicts the electron density in the O$^{\geq 2+}$ region to be 10$^4$ ${\leq}$ n${_e} ($cm$^{-3}$)${\leq}$ 7${\times}$10$^5$, which although is below the critical density for collisional de-excitation of the ${^1D_2}$ level responsible for producing [O\,{\sc iii}]${\lambda\lambda}$4959,5007 forbidden emission lines, is significantly higher
than the 1.5${\times}$10$^4$ cm$^{-3}$ critical density for the transition responsible for the 25.9 ${\mu}$m [O\,{\sc iv}] emission line. Thus, Cloudy predicts the [O\,{\sc iv}] emission 
line shape and intensity will be affected by gas density, but the same is not true for the [O\,{\sc iii}]${\lambda\lambda}$4959,5007 forbidden emission lines. 

The ionization structure depends very
sensitively on the density of the inflowing neutral H gas. Figure 3 illustrates the dependence in an
animation consisting of a series of graphs for which the neutral H density at the reverberation radius;
3.8 lt-days or 3.5${\times}$10$^{-3}$ pc \citep{Den10}, is progressively increased from 7.1 ${\leq}$ hden ${\leq}$ 8.3 dex. 
Systematic trends are apparent, and are summarized as follows. First, as the neutral H density increases, the size of the H$^{+}$ emitting region decreases as expected due to increased photoionization opacity.  Indeed, when the neutral H gas density at the reverberation radius exceeds 2 ${\times}$ 10$^8$ cm$^{-3}$ the central UV--X-ray source is unable to photoionize an H nebula with any appreciable radial extent at all, and hence is unable to produce any broad H emission lines. Second, as the neutral H density increases, the O$^{\geq 2+}$ and C$^{\geq 2+}$
regions move progressively closer to the central UV--X-ray source, and diminish in size. Thirdly, as the neutral H density increases,
the O$^{+}$ and O$^{0}$ regions move closer to the central UV--X-ray source which could result in broad vacuum wavelength [O\,{\sc i}]${\lambda\lambda}$6302,6365 emission lines. The fact that broad [O\,{\sc i}] emission lines are not observed (see Figure 1), 
even though the density in the O$^{0}$ region is below the critical density, 9 ${\times}$ 10$^{5}$ cm$^{-3}$, allows a firm constraint to be placed on the neutral H density for the inflowing gas; hden ${\leq}$ 7.3 dex at the reverberation radius. Since ionized H dominates the electron density, ${n_e}$ ${\leq}$ 7.3 dex at the reverberation radius also. At larger radii, ${n_e}$ asymptotes to ${\sim}$3.65 dex, close to the observed value measured with the [S\,{\sc ii}] ${\lambda}$6742/${\lambda}$6757 ratio which corresponds to ${n_e}$ ${\sim}$ 10$^3$ cm$^{-3}$, at a radial distance of ${\sim}$0.5 pc. Interestingly, the C\,{\sc iii}]${\lambda}$1908 line is a doublet, sensitive to gas density \citep{Kee92}. In principle, the {\it Cosmic Origins Spectrograph} aboard ${\it HST}$ could resolve the C\,{\sc iii}] doublet in NGC 3227 providing an independent measure of the electron density inside the O$^{\geq 2+}$ region because the O$^{\geq 2+}$ and C$^{\geq 2+}$ regions are predicted to be co-extensive (see Figure 3). 

\subsection{Emission Line Ratios}

What is visually striking about the spectra obtained with ${\it STIS}$ of NGC  3227 is how weak the 
emission lines actually are compared to the H Balmer emission lines, especially the vacuum wavelength [O\,{\sc ii}]${\lambda\lambda}$3727,3729 lines which 
rendered NGC 3227 unclassifiable according to the diagnostic diagram of \cite{Kew06}, as noted previously by \cite{D13}. Emission line ratios are a particularly sensitive diagnostic of the physical conditions in the ionized gas including the gas density, temperature and metallicity. Table 3 presents the observed emission line fluxes, normalized to H${\beta}$, for multiple transitions of H, C, N, O,  and S.
When one compares the observed emission line ratios to the intrinsic values\footnotemark \footnotetext[2]{employing the {\it aperture slit} option within Cloudy}  predicted by the photoionization code Cloudy, the predicted emission line ratios for C, N, and O, are about one order of magnitude brighter than observed as shown in Table 3, and illustrated in Figure 4. The lines in question cover the gamut in wavelength, but the nature of the discrepancies are such that they can not be attributed solely to dust extinction.
The differences are most apparent for the C, N and O lines, suggesting a metallicity
dependence. Interestingly, as Figure 4 illustrates, the discrepancy between the observed and predicted
values diminishes when the chemical abundance of C, N and O are decreased by a factor of 
ten compared to the ISM values quoted in the Cloudy documentation. Evidently, one explanation for the reason why the C, N and O emission lines are observed to be so faint in the AGN of NGC 3227 is because the gas there is metal poor, suggesting an origin in the circumgalactic medium. Perhaps the inflow, that we perceive as the BLR in NGC 3227, is just the terminus of a much larger inflow that originates from outside the galaxy. Such inflows of metal poor gas appear to be commonplace, observed in our own Galaxy \citep{JBH07} and others \citep{L13}, but this is perhaps the first association 
between the BLR of an AGN and a low metallicity accretion flow from the circumgalactic medium. In the case of NGC 3227, the inflow may be facilitated by the interaction with NGC 3226 \citep{Mun95,Mun96}.

\begin{deluxetable}{cccc}
\tabletypesize{\scriptsize}
\tablecaption{Comparing Emission Line Ratios}
\tablewidth{0pt}
\tablehead{
\colhead{Line\tablenotemark{a}} & \colhead{Observed Ratio\tablenotemark{b}} & \colhead{Cloudy\tablenotemark{c}}  & \colhead{Cloudy\tablenotemark{c}} \\
\colhead{} & \colhead{} & \colhead{C,N,O = 0.1 ISM}   & \colhead{C,N,O = ISM}\\
\colhead{(1)} & \colhead{(2)} & \colhead{(3)} & \colhead{(4)}    \\
}
\startdata
$\textrm{[S\,{\sc ii}] 6732}$ &  0.05 & 0.07 &  0.08 \\
$\textrm{[S\,{\sc ii}] 6718}$ &  0.04 & 0.04  &  0.05 \\
$\textrm{[N\,{\sc ii}] 6585}$ &  0.09 & 0.07 &  1.0 \\
H${\alpha}$  6564& 4.1 & 3.75  &  4.2 \\
$\textrm{[N\,{\sc ii}] 6550}$ &  0.03 & 0.02 &  0.35 \\
$\textrm{[O\,{\sc i}]}$ 6365 & 0.01 & 0.02  & 0.2 \\
$\textrm{[O\,{\sc i}]}$ 6302 & 0.02 & 0.06  & 0.8 \\
$\textrm{[O\,{\sc iii}]}$ 5008 & 0.77 & 0.84 &  3.9 \\
$\textrm{[O\,{\sc iii}]}$ 4960 & 0.24 & 0.28  & 1.3 \\
$\textrm{[O\,{\sc iii}]}$ 4364 & ${\leq}$ 0.02 & 0.14 &  0.52 \\
H${\beta}$ 4862 & 1 & 1 & 1 \\
H${\gamma}$ 4341 & 0.3 & 0.46 & 0.47 \\
$\textrm{[O\,{\sc ii}]}$ 3727,3729 & ${\leq}$ 0.01 & 0.02 & 0.43 \\
$\textrm{ Mg II}$ 2798 & 0.23 & 0.01  & 0.01 \\
$\textrm{C III]}$ 1908 & 0.18 &  3.4 & 15.3 \\
$\textrm{He II}$ 1640 & 0.03 &  5.23 & 4.46 \\
$\textrm{C IV}$ 1542 & ${\geq}$0.15 & 1.1 & 1.76 \\
\enddata
\tablenotetext{a}{Vacuuum wavelength.}
\tablenotetext{b}{Relative to H${\beta}$. Uncertainties in the observed ratio are ${\sim}$10\%. Observed fluxes are measured within
a 0.2{\arcsec}  x 0.35{\arcsec}  aperture. They are continuum subtracted but not corrected for dust extinction. Model dependent systematic uncertainties introduce an additional ${\sim}$3\% uncertainty not reported in the Table. }
\tablenotetext{c}{Intrinsic values employing the aperture slit option within Cloudy.}
\end{deluxetable}

\begin{figure}
\epsscale{1.0}
\begin{center}
\includegraphics[width=84mm,clip]{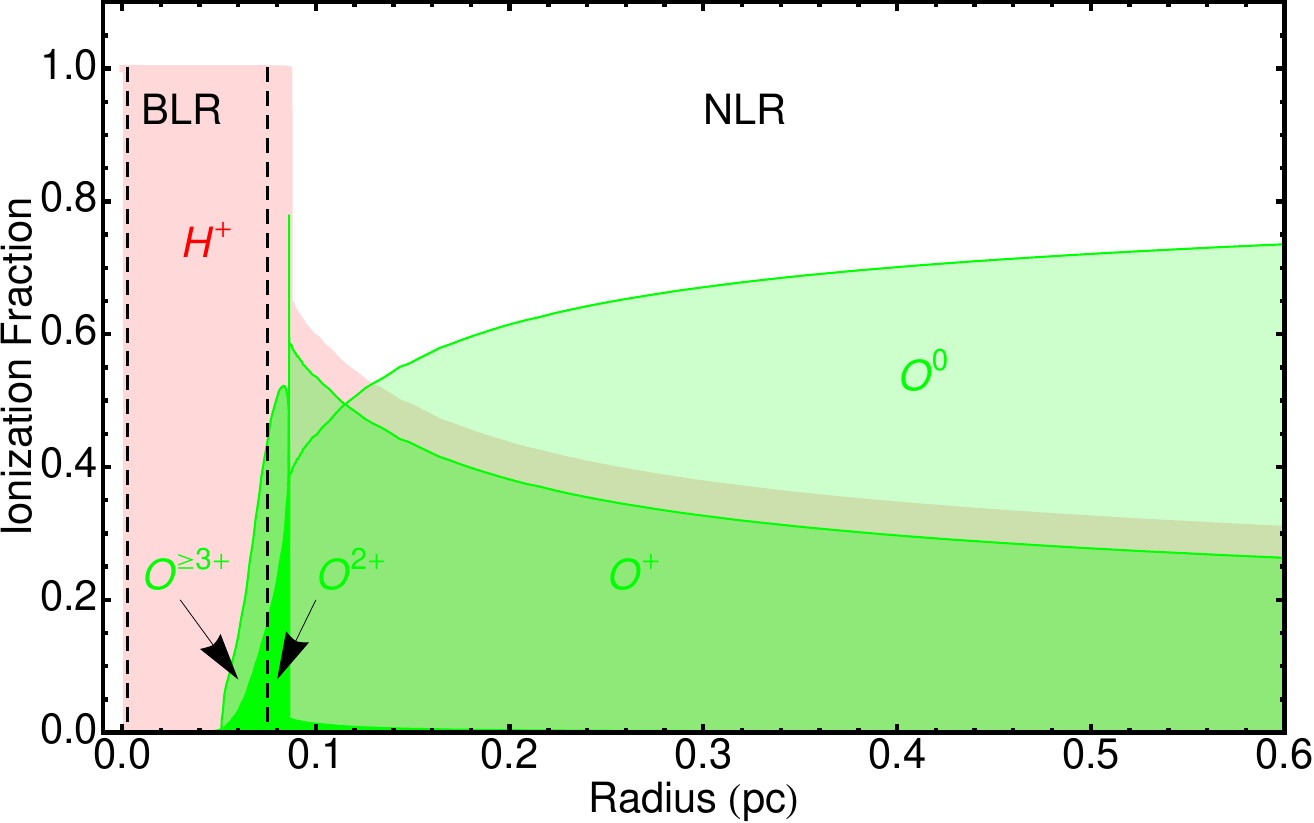}
\caption{{Photoionization model for NGC 3227. The radial dependence of the ionization fraction for H (pink shading), and various O ions (green shading) is shown in the top left panel. The radial dependence of the ionization fraction for H (pink shading), and various C ions (blue shading) is shown in the top right panel. The electron number density is shown in the lower left panel. The data point with the error bar represents the electron density measured
using the  [S\,{\sc ii}] ${\lambda}$6742/${\lambda}$6757 ratio which corresponds to ${n_e}$ ${\sim}$ 10$^3$ cm$^{-3}$, at a radial distance of ${\sim}$0.5 pc.  The neutral H number density, ${\propto r^{-3/2}}$ is shown in the lower right panel. Vertical dashed lines identify the inner, $r{_i}$, and outer radii, $r{_o}$, of the BLR in units of pc \citep{D13}. (An animation and a color version of this figure are available in the online
journal.) }}
\label{default}  
\end{center}
\end{figure}

\begin{figure}
\epsscale{1.0}
\begin{center}
\includegraphics[width=84mm,clip]{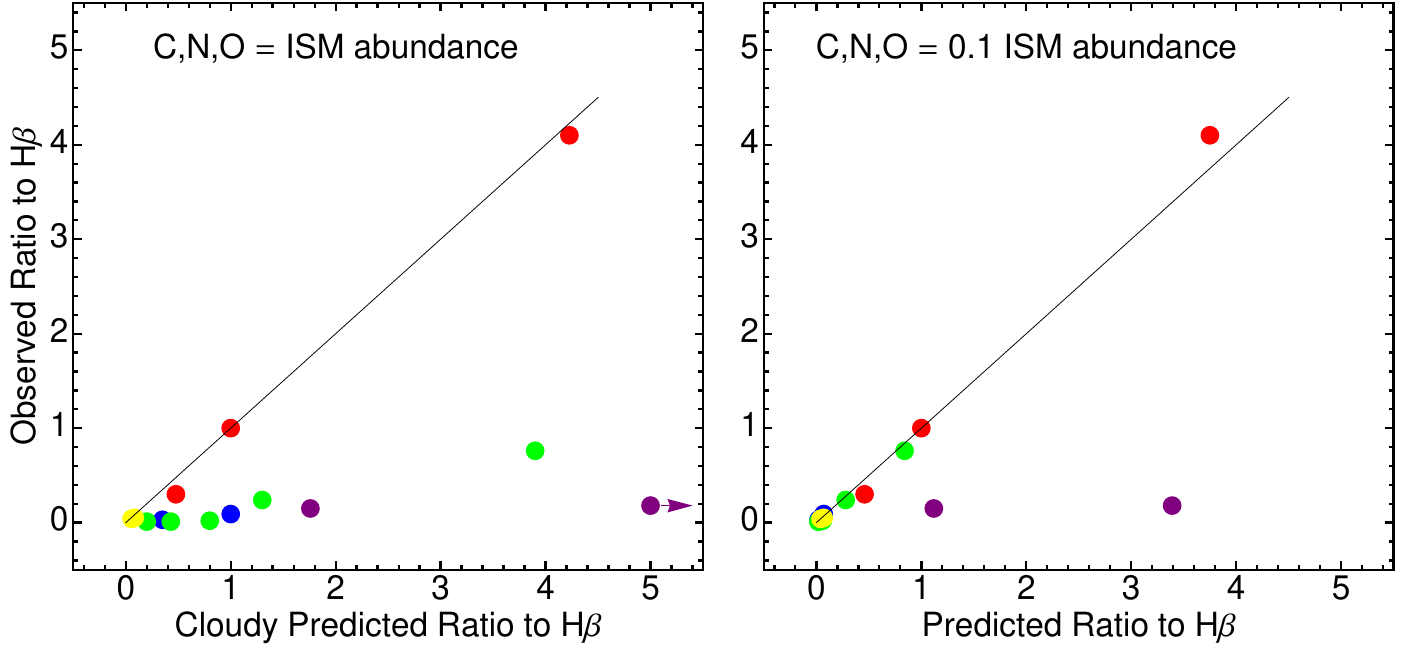}
\caption{{Comparing emission line ratios, normalized to H${\beta}$, for H (red), C (purple), N (blue), O (green) and S (Yellow) observed with ${\it STIS}$ on the ordinate and those predicted by Cloudy on the abscissa (see Table 3). The left hand panel shows the correlation for emission line ratios predicted by Cloudy using ISM abundances. The right hand panel shows the correlation with emission line ratios predicted by Cloudy for gas that has been depleted in C, N, and O, by a factor of ten. }}
\label{default}  
\end{center}
\end{figure}

\vspace{1.0cm}
\subsection{UV lines and Dust Extinction}

The fact that the emission lines observed with {\it STIS} span such a large range in wavelength provides an opportunity to compare the dust extinction inferred from the emission lines with a prior study by \cite{Cre01}
that employed the continuum. As Table 3 and Figure 4 illustrate, the good agreement between the observed and predicted emission line ratios for the low metallicity gas suggests that the dust extinction to the visible lines is virtually negligible. The only exceptions are the vacuum wavelength C\,{\sc iii}]${\lambda}$1908 and C\,{\sc iv}${\lambda}$1542 emission lines.
Comparing the observed C\,{\sc iv}/H${\beta}$ emission line ratio with the intrinsic one predicted by Cloudy for the metal poor ISM leads to 
a differential extinction, A$_{1542}$ - A$_{H{\beta}}$ = 2.2 mag which is lower than the 3.8 mag estimated
using the extinction curve for NGC 3227 presented by \cite{Cre01}. Indeed, the inferred dust extinction could be a further 0.75 mag smaller if one compensates for the fact that the C\,{\sc iv} emission line suffers from intrinsic absorption \citep{Cre01}.  On the other hand the differential extinction, A$_{1908}$ - A$_{H{\beta}}$, estimated for the C\,{\sc iii}]/H${\beta}$ emission line ratio is 3.2 mag
which is higher than the 2.5 mag estimated using the extinction curve for NGC 3227 presented by \cite{Cre01}. 
Taken together, the differences between the observed  C\,{\sc iii}], and C\,{\sc iv} line ratios, and those predicted by Cloudy, are inconsistent with a uniform screen of dust obscuration. 

An additional puzzle involves the vacuum wavelength Mg\,{\sc ii} ${\lambda}$2798 emission line which is observed to be about a factor of 20 brighter than predicted by Cloudy, a situation that is only exacerbated by employing an extinction correction.
However, the most remarkable discrepancy lies with the He\,{\sc ii} ${\lambda}$1640 line which is observed to be
an order of magnitude fainter than predicted by Cloudy, even after correcting the observed line for a plausible 3 mag of extinction. The ions responsible for producing the Mg\,{\sc ii} ${\lambda}$2798 and He\,{\sc ii} ${\lambda}$1640 emission lines require photons with energies exceeding 7 eV, and 54 eV respectively. That these lines represent the extrema of ionization potentials sampled with the ${STIS}$ spectra suggests that a remedy lies with the spectral shape adopted for the ionizing continuum. However, the discrepancy between the observed and predicted values for Mg\,{\sc ii} ${\lambda}$2798 and He\,{\sc ii} ${\lambda} $1640, persists over a range in spectral index; -1.4 ${\leq}$ ${\alpha_{\rm{ox}}}$ ${\leq}$-1.0, and blackbody temperatures; 10$^4$ ${\leq}$ T(K) ${\leq}$ 3${\times}$10$^{5}$ selected for the Cloudy models.

The UV spectrum of NGC 3227 is also unusual when compared to other AGNs. For example, the C\,{\sc iv} ${\lambda}$1542 emission line is much brighter than the adjacent C\,{\sc iii}] ${\lambda}$1908 emission line in the Sy 1 NGC 5548 \citep{K20}, and in nearby quasars \citep{N13}. Whereas in NGC 3227, the C\,{\sc iv} ${\lambda}$1542 and C\,{\sc iii}] ${\lambda}$1908 emission lines are observed to be of comparable brightness. 

In summary, the Cloudy photoionization model can succesfully explain the relative intensities of the H, N, O and S emission lines seen in the visible, in terms of a low metallicity inflow photoionized by the central UV-X-ray source. However, the same model is unable to simultaneously explain the UV emission lines, especially the bright Mg\,{\sc ii} ${\lambda}$2798 line, and the faint He\,{\sc ii} ${\lambda}$1640 line. 

\subsection{Comparison with the Standard Broad Line Region Model}

The main feature of the model presented here for NGC 3227 is photoionization of low density gas ${\leq}$ 10$^7$ cm$^{-3}$ which leads to a spatially extended spherically symmetric nebula surrounding the central UV--X-ray source. In contrast, the so called `standard model' for the BLR in Sy 1 nuclei and quasars refers to photoionization of high density gas; 10$^7$ cm$^{-3}$ ${\leq}$ n${_e}$ ${\leq}$ 10$^{14}$ cm$^{-3}$   \cite[e.g.,][]{KK81,F92,K20,N13}. The main reason why the `standard model' would fail to reproduce the UV--visible spectrum of NGC 3227 is because such high gas densities would quench the ionizing photons so completely that there would be no radial extent to the ionized nebula, and hence no broad emission lines, and no ionization gradient.  As noted previously by \cite{K84}, a viable photoionization model must explain not only the emission line ratios but also the emission line shapes.
As yet, there is no computer code that incorporates both photoionization and kinematics in such a way as to predict emission line shapes, but the development of such a capability is clearly the way forward to a complete understanding of the emission line spectra of AGNs, particularly given the wealth of information 
gathered with {\it STIS}. 

In the case of NGC 3227, the broad emission lines allow the ionization structure of 
the photoionized nebula surrounding the AGN to be deciphered, empirically, by comparing the emission line profiles in velocity space as illustrated in Figure 5. The H${\alpha}$ profile defines the extent of the BLR in velocity space. The broad wings of the H${\alpha}$
line match those of the C\,{\sc iv} profile. The two profiles depart at velocities with an absolute value less than 1000 km/s due to intrinsic absorption of the C\,{\sc iv} line, attributed to foreground, low density, gas in the disk of this highly inclined galaxy \citep{Cre01}. However, the core of the C\,{\sc iv} line is not important here, only the broad wings which the [O\,{\sc iii}] line does not have. By contrast with the C\,{\sc iv} line,
the [O\,{\sc iii}] line is narrow with no obvious broad component. Photoionization modeling with Cloudy suggests that the electron density in the O$^{\geq 2+}$ region is below the critical density for collisional de-excitation of the ${^1D_2}$ level responsible for producing [O\,{\sc iii}] ${\lambda}$5007 forbidden emission line (see section 3.1). Consequently,  if one accepts that the C\,{\sc iv} line provides a surrogate measure of the extent, in velocity space, of high ionization photons with energies ${\ge}$ 48 eV, then the reason why the broad wings of the C\,{\sc iv} line coincide with diminished [O\,{\sc iii}] line emission, is because O$^{2+}$ is progressively ionized to O$^{3+}$ as the central UV--X-ray source is approached. Such an ionization gradient is predicted by Cloudy (see Figure 3) and observed with {\it STIS} (see Figure 1). 

\begin{figure}
 \includegraphics[width=84mm,clip]{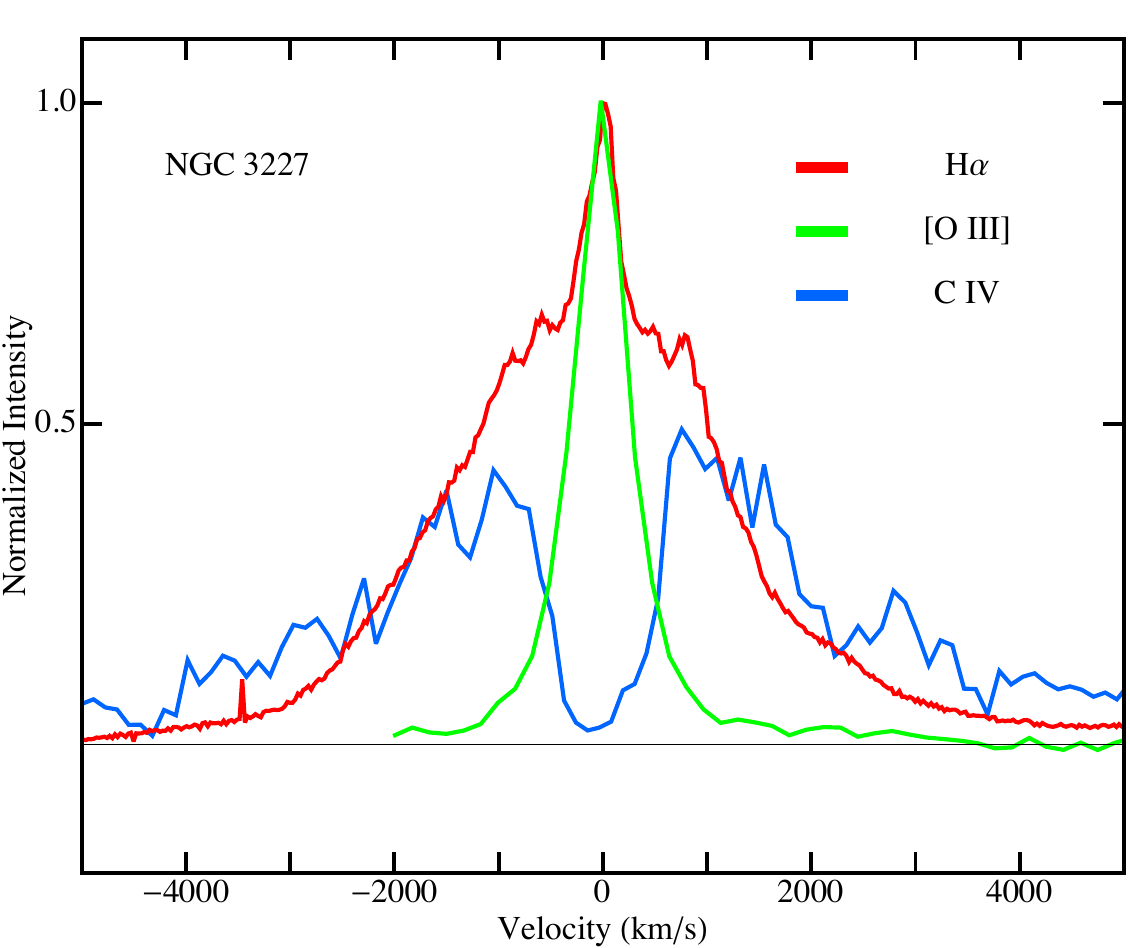}
 \caption{{\it HST} spectra showing a detailed comparison of the H${\alpha}$ (red),  [O\,{\sc iii}] (green) and C\,{\sc iv} (blue) emission line profile shapes with their respective wavelengths converted to velocity using the non-relativistic Doppler equation. The H${\alpha}$ and [O\,{\sc iii}] profiles are normalized to their respective peak intensity. The C\,{\sc iv} profile is normalized, arbitrarily, to match the H${\alpha}$ profile at ${\pm}$ 1300 km/s.  }
\end{figure}

\section{Conclusion}

This {\it Paper} presents evidence challenging the conventional wisdom that the absence of broad [O\,{\sc iii}] forbidden emission lines implies that the gas density is high enough, ${n_e}$ ${\ge}$ 7 ${\times}$ 10$^5$ cm$^{-3}$, to cause collisional de-excitation of the ${^1D_2}$ level of O$^{2+}$ in the BLR of AGNs. On the contrary, the absence of broad [O\,{\sc iii}] forbidden emission lines is attributed to photoionization of O$^{2+}$, and not high density, at least in NGC 3227. The observational evidence includes an ionization gradient 
defined by a rich spectrum of UV--visible emission lines attributable to H, He, C, N, O, Mg, and S. Additionally, a comparison of emission line profile shapes reveals that high velocity permitted C\,{\sc iv} ${\lambda}$1549 line emission coincides with diminished [O\,{\sc iii}] ${\lambda}$5007 forbidden line emission as would be expected if O$^{2+}$ is progressively ionized to O$^{3+}$. These observational results 
constrain a model in which the central UV-X-ray source in NGC 3227 photoionizes a steady-state inflow of low metallicity gas.

\section{Acknowledgments}

{\it Facilities:}  \facility{HST (STIS)}


\begin{thebibliography}{99}


\bibitem[Bland-Hawthorn et al. (2007)]{JBH07}  Bland-Hawthorn, J., Sutherland, R., Agertz, O., \& Moore, B., 2007, apjl, 670, 109

\bibitem[Crenshaw et al. (2001)]{Cre01} Crenshaw D.M., Kraemer S.B., Bruhweiler F.C., Ruiz J.R., 2001, ApJ 555, 633	

\bibitem[\protect\citeauthoryear{Dasyra et al.}{2011}]{D11} Dasyra K. M., Ho L. C., Netzer H., Combes F., Trakhtenbrot B., Sturm, E., Armus L., Elbaz D., 2011, ApJ 740, 94

\bibitem[Denney et al. (2010)]{Den10} Denney K.D., et al., 2010, ApJ 721, 715

\bibitem[\protect\citeauthoryear{Devereux}{2013}]{D13} Devereux N., 
2013, ApJ 764, 79

\bibitem[\protect\citeauthoryear{Ferland et al.}{1992}]{F92} Ferland G. J., Peterson B. M., Horne K., Welsh W. F. \& Nahar S. N., 1992, ApJ 387, 95

\bibitem[\protect\citeauthoryear{Ferland et al.}{2013}]{F13}  Ferland, G. J., Porter, R. L.,van Hoof, P. A. M., Williams, R. J. R., Abel, N. P., Lykins, M. L., Shaw, G., 
Henney, W. J., and Stancil, P. C. The 2013 Release of Cloudy. Revista Mexicana de Astronomia y
Astrofisica, 49, 137.

\bibitem[\protect\citeauthoryear{Khachikyan \& Weedman}{1971}]{KW71} Khachikyan E.Y.,  Weedman D. W., 1971, Ap 7, 231

\bibitem[Kewley et al. (2006)]{Kew06} Kewley, L.J., Groves, B., Kauffmann, G., \& Heckman, T., 2006, \mnras, 372, 961

\bibitem[Keenan, Feibelman, \& Berrington (1992)]{Kee92} Keenan, F. P., Feibelman, W. A., \& Berrington, K. A., 1992, \apj, 389, 443

\bibitem[\protect\citeauthoryear{Korista \& Goad}{2000}]{K20}	Korista K. T.,  Goad M. R., 2000, ApJ 536, 284
	
\bibitem[\protect\citeauthoryear{Kwan}{1984}]{K84} Kwan J., 1984, ApJ 283, 70

\bibitem[\protect\citeauthoryear{Kwan \& Krolik}{1981}]{KK81} Kwan J., Krolik J. H., 1981, ApJ 250, 478

\bibitem[\protect\citeauthoryear{Lehner et al.}{2013}]{L13} Lehner, N.,  Howk, J. C.,  Tripp, T. M.,  Tumlinson, J.,  Prochaska, J. X., O'Meara, J. M., Thom, C., Werk, J. K., Fox, A. J., Ribaudo, J., \apj, 770, 138
	
\bibitem[Mundell et al. (1995)]{Mun95}  Mundell, C. G., Pedlar, A., Axon, D. J., Meaburn, J., \& Unger, S. W., 1995, \mnras, 277, 641

\bibitem[Mundell et al. (1996)]{Mun96}  Mundell, C. G., Pedlar, A., Shone, D. L., Axon, D. J., Meaburn, J., \& Unger, S. W., in Barred Galaxies, ASP Conference Series, Vol. 91, p.473

\bibitem[\protect\citeauthoryear{Negrete et al.}{2013}]{N13} Negrete C. A.,  Dultzin D., Marziani P., \& Sulentic J. W., 2013, ApJ 771, 31

\bibitem[Osterbrock \& Ferland (2006)]{Ost06} Osterbrock D., \& Ferland G. J., 2006, Astrophysics of Gaseous Nebulae and Active Galactic Nuclei, 2nd ed. University Science Books, Sausalito, CA. 

\bibitem[Rubin (1989)]{Rub89} Rubin R.H., 1989, ApJS 69, 897

\bibitem[\protect\citeauthoryear{Shields, Oke \& Sargent}{1972}]{S72} Shields G. A., Oke J. B., Sargent W. L. W., 
1972, ApJ 176, 75

\bibitem[\protect\citeauthoryear{Sulentic et al.}{2007}]{S07} Sulentic J. W., Bachev R., Marziani P., Negrete C. A., Dultzin D., 2007, ApJ 666, 757

\bibitem[Vasudevan \& Fabian (2009)]{Vas09} Vasudevan, R.V., \& Fabian, A.C., 2009, \mnras, 392, 1124

\bibitem[Winter, Veilleux, McKernan \& Kallman (2011)]{Win11} Winter L.M., Veilleux S.,  McKernan B., Kallman, T.R., 2011, ApJ 745, 107

\end{thebibliography}
\end{document}